\documentclass[12pt]{article}

\usepackage{multibib}
\newcites{SI}{References for Supporting Information}
\usepackage[colorlinks=true,linkcolor=blue,citecolor=blue]{hyperref}
\usepackage{PRIMEarxiv}
\usepackage[utf8]{inputenc}
\usepackage[T1]{fontenc}
\usepackage{cmbright}
\usepackage{hyperref}
\usepackage{url}
\usepackage{booktabs}
\usepackage{amsfonts}
\usepackage{nicefrac}
\usepackage{microtype}
\usepackage{lipsum}
\usepackage{fancyhdr}
\usepackage{graphicx}
\usepackage{ragged2e}
\usepackage[table]{xcolor}
\usepackage{float}
\usepackage{caption}
\graphicspath{{media/}}
\usepackage{amsmath}
\usepackage{cite}
\usepackage{hyperref}

\title{PSEUDOPERIODIC SPHERICAL BOUNDARY CONDITIONS: EFFICIENT AND ISOTROPIC 3D PARTICLE SIMULATIONS WITHOUT LATTICE ARTIFACTS}
\author{
  \And
 \textbf{MANUEL DEDOLA$^{\dag}$, LUDOVICO CADEMARTIRI$^{\dag *}$} \\
$^{\dag}$Department of Chemistry, Life Sciences and Environmental Sustainability, University of Parma,\\
Parco Area delle Scienze 17 A, Parma, Italy. \\
$^{*}$Author to whom correspondence should be addressed: \\
ludovico.cademartiri@unipr.it 
}

\begin{document}
\maketitle

\RaggedRight 

\begin{abstract}

Periodic Boundary Conditions (PBC) introduce well-known lattice artifacts. We present a novel Pseudoperiodic Spherical Boundary Condition (SBC) that is perfectly isotropic. Through detailed comparative simulations, we demonstrate that SBC eliminates the structural and dynamic anisotropy inherent to PBC. For the crowded systems where these artifacts are most prominent, our method is also computationally more efficient than standard Minimum Image Convention implementations. This establishes SBC as a powerful, high-fidelity alternative for simulating isotropic matter.
\end{abstract}

\section{Introduction}

Many of the most scientifically and technologically important systems in nature are crowded, starting from the cellular medium.\cite{1,2,3,4,5} Arguably, the characterizing traits of these systems are two: (i) their emergent (and sometimes chaotic) behaviour\cite{6,7}, especially when driven outside of equilibrium, and (ii) their notorious mechanistic intractability.
Simulating these dense environments is a profound challenge\cite{8} and not only for the obvious reasons (e.g., complex, emergent phenomena such as caging, subdiffusion, and cooperative molecular rearrangements that test the limits of mean-field theories\cite{9,10}). Crowded systems have the unfortunate characteristic of amplifying the influence of simulation artifacts, especially as the simulations times increase\cite{11,12,13,14}. It is therefore important to identify and try to circumvent all sources of artifacts.
Among the most important sources of artifacts are the boundary conditions. Crowded systems are most often simulated using Periodic Boundary Conditions (PBCs)\cite{15}, particularly when long-range interactions are present\cite{16}. PBCs have three main advantages. (i) They eliminate physical surfaces. The boundary is permeable: particles that exit one face of the unit cell re-enter from the opposite face. This fact conserves the mass in the simulation volume and allows for the simulation of bulk systems without the severest artifacts caused by hard-wall containers (e.g., particle layering, density gradients, reflections at the boundary)\cite{15}. (ii) They provide efficient “tricks” to model interactions with particles outside the simulation volume. For short-range interactions, “ghost particles” – symmetrically equivalent copies of particles near a boundary – can be generated in the adjacent image cells to correctly compute forces. A more common alternative is the Minimum Image Convention (MIC), a mathematical rule by which particles interact only with the closest periodic image of any other particle. Non-truncated long-range forces, such as electrostatics, can instead be estimated by solving part of the problem in Fourier space, leveraging the system’s periodicity\cite{17}. (iii) They drastically improve computational efficiency. By accurately modeling a pseudo-infinite bulk environment, PBCs allow for the robust calculation of thermodynamic and structural properties using what is usually a computationally tractable number of particles\cite{15}.
Unfortunately, as they say, “there is no such thing as a free lunch”, and PBCs are responsible for a host of issues. On the computational side, explicit ghost particle methods introduce memory and particle overhead (for every particle within a cutoff distance of a boundary, up to 7 ghost copies must be created and stored) and computational complexity (e.g., interactions must be computed for this expanded set of real and ghost particles)\cite{18}. MIC, on the other hand, relies on a per-pair calculation (the distance-wrapping operation must be applied on each interaction that crosses a boundary, which can be less efficient than a distance calculation)\cite{19,20}, encumbers communication in parallel implementations (with decomposed domains, processors must communicate boundary-zone particle data to their neighbors, creating communication latency)\cite{18,19,20,21}, is problematic in its point-wise application to large molecules or aggregates, and scales poorly with large objects (to prevent ripping them apart, the simulation box must be much larger than the object itself, which is highly inefficient for studying large or aggregating systems as it requires simulating vast amounts of empty space)\cite{15}.\\

The physics issues are even more significant:\\

\begin{enumerate}
    \item Finite-size effects. The finite box size artificially and anisotropically truncates the spectrum and the intensity of density and energy fluctuations, which causes systematic biases in the calculation of thermodynamic response functions like compressibility and heat capacity\cite{22,23,24}. 
    \item Artificial periodicity effects. (a) Artificial recurrence: a particle's trajectory is unnaturally folded back into the primary cell, creating spurious, short-path returns that alter long-timescale diffusion and kinetic rates\cite{25,26}. (b) Violating conservation laws: the standard particle-wrapping algorithm discontinuously changes particle position vectors, thus violating the conservation of total angular momentum, which is critical for systems involving rotation or shear\cite{27}. (c) Incompatibility with interfaces: simulating inhomogeneous systems like droplets or flat interfaces requires computationally wasteful vacuum slabs or introduces unphysical interactions with periodic images, distorting the system’s intrinsic properties\cite{28}.
    \item Boundary shape effects. The fundamental mismatch between the imposed cubic symmetry of the PBC lattice and the isotropic nature of inter-particle potentials introduces a pervasive structural bias\cite{29}. The cubic geometry “trickles down” into an artificial orientational order on the fluid, most critically in the nearest-neighbor shell. This structural bias extends to the energetics via Ewald summation and results in anisotropic dynamics, where quantities like the diffusion coefficient become tensors instead of scalars. Consequently, thermodynamic properties derived from this system are not those of a true isotropic fluid, but of one artificially and non-trivially ordered by its container\cite{30}.
\end{enumerate}

\begin{figure}
	\centering
	\includegraphics[width=0.5\linewidth]{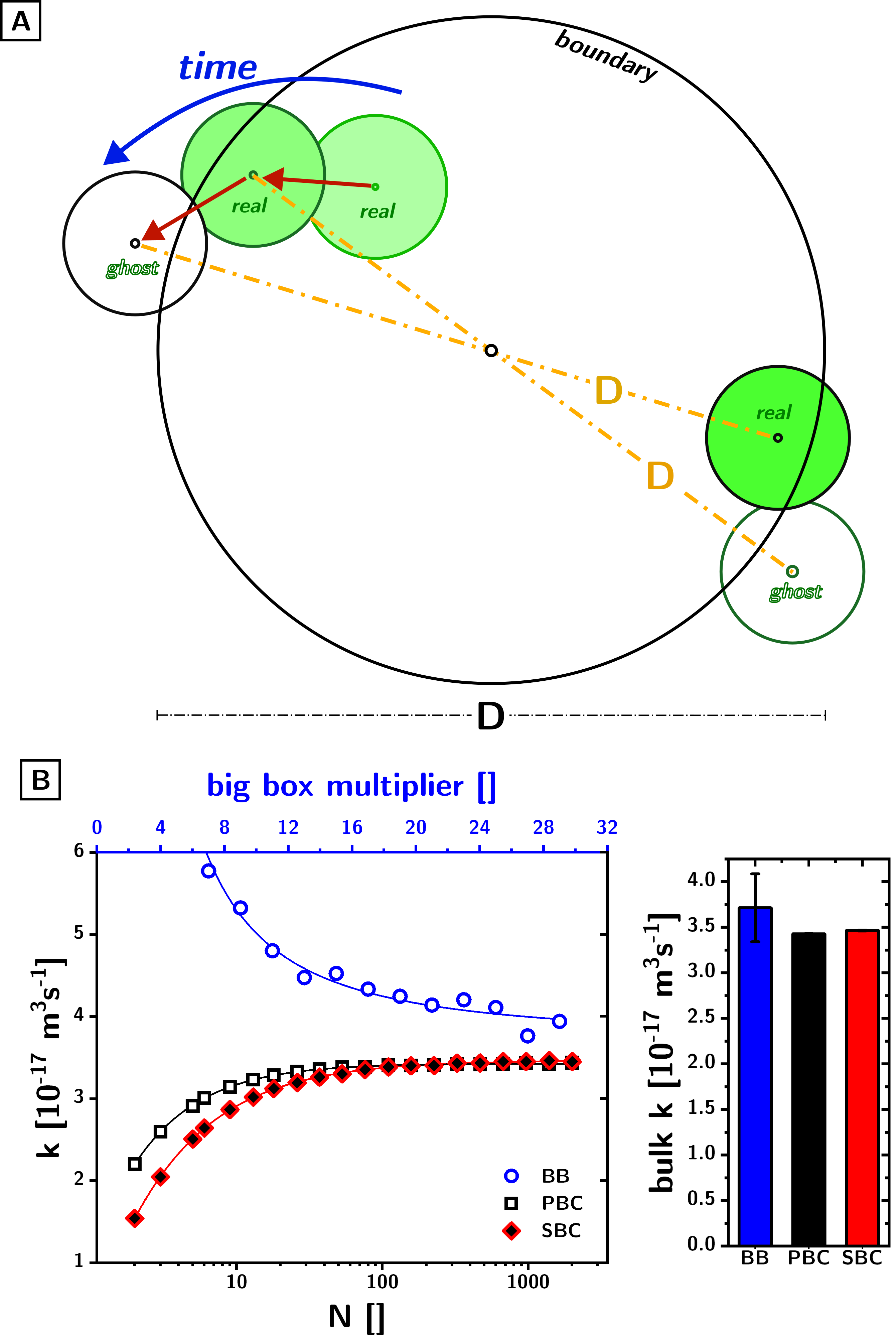}
	\caption {\textbf{Spherical boundary conditions (SBCs) and their convergence}. \textbf{(A)} Sketch of the boundary rules in SBCs with single ghost particle generation in the antipodal direction by a simulation volume diameter. \textbf{(B)} Convergence of collision rate for SBC and PBC as a function of N compared to the convergence of a central-region analysis in a hard-walled cubic box (“Big Box”, BB) of increasing size. On the right, the extrapolated asymptote values of the three boundary conditions.}
    \label{Fig1}
\end{figure}

While various techniques exist to mitigate these issues (e.g., finite-size scaling\cite{31}, Lees-Edwards boundaries\cite{32} and many others\cite{33,34,35,36,37,38,39,40,41,42,43,44,45,46}, they add significant complexity and are often imperfect approximations that do not cure the underlying problems.
In light of these persistent challenges we wondered whether it was possible to preserve most of the key assets of PBCs (i.e., permeable/soft boundary), while doing away with its most troubling issues caused by its anisotropy and forced periodicity. We propose therefore Spherical Boundary Conditions (SBCs) designed to simulate isotropic systems, at this stage particularly for Brownian Dynamics of crowded systems, as a potentially superior alternative to PBC in certain contexts.
SBCs are defined as follows (cf. Figure \ref{Fig1}A). Objects are simulated within a primary spherical volume of radius R. When any part of an object enters a boundary shell of thickness $r_c$ (which can be the interaction threshold distance, e.g., $r_c$ = radius of the particle for hard sphere potentials), a “ghost copy” of the entire object is created. The ghost is placed with its center of mass at a distance 2R from that of its real counterpart, along the antipodal direction, without rotations or reflections. The real object is deleted and the ghost is promoted to “real” status upon the real object’s complete exit from the sphere, preserving the kinematic state. The direct interaction between a real particle and its own ghost can be explicitly turned off to prevent a g(r) artifact at r=2R.
The goal of this work was then to explore the characteristics of these SBCs, especially in comparison with cubic SBCs, in terms of dynamics, structure, correlations and computational performance.

\section{Computational Design}

For this initial validation of our boundary condition, we employed spherical particles, hard-sphere potentials, and Brownian Dynamics (BD). This combination offers a clear and computationally efficient framework to isolate the effects of the boundary conditions themselves.\\

\textbf{Spherical particles}. They provide analytical tractability to boundary crossings, collisions, and the determination of a rigorous Brownian trajectory.\\

\textbf{Hard-sphere potentials (HSP)}. Pragmatically, hard sphere potentials simplify our implementation and make the code more transparent. Methodologically, because PBC’s artifacts - like long range structural correlations - are muted for hard spheres, they offer a stringent testbed for exploring alternative boundary rules\cite{35}. Also, HSPs make MIC and explicit ghost particles approaches for PBCs analogous in terms of structure.\\

\textbf{Brownian dynamics (BD)}. BD avoids the computational overhead of simulating explicit solvent molecules, which allows - as opposed to traditional Molecular Dynamics (MD) - for those substantial simulation times required to probe the long-time structural correlations induced by the boundary geometry and to ensure the thorough equilibration of the crowded systems at the core of this work. Given its widespread use and its ability to capture the essential physics of diffusion and caging, BD represents a highly relevant and computationally tractable framework for this study.\\

As previously demonstrated the accurate description of a Brownian trajectory by a memoryless random walk requires using timesteps in the order of $200\tau_r$ (where $\tau_r$ is the relaxation time) and the correction of the diffusivity\cite{47}. The $\tau_r$ was determined by Green-Kubo integration of the Velocity Autocorrelation Function (VACF) for spheres in fluids that accounts for hydrodynamic and acoustic autocorrelation effects\cite{48,49,50,51,52,53,54}. The correction factors $\alpha$ on the diffusivity were obtained by the subsampled VACF-determined physical trajectories (cf. Supporting Information). Unless stated otherwise our systems were composed of N=1000 $SiO_2$ particles, with a radius r=1.12 nm, in water at 298.15K for which $\alpha = 0.7074$.\\
For PBC boundary conditions we employed an explicit ghost particle method rather than MIC, in order to isolate the influence of the boundary on the structure and dynamics. Of course, when we compared SBCs computational performance to PBCs we employed MIC as they are state of the art.\\

\textbf{Collision kinetics}. The collective observable in a particle system that is most easily and unambiguously measureable is the collision rate. As such it is the ideal testbed to compare the finite size effects in SBCs and PBCs and ensure they both reach the same limit behavior as N increases. The ground truth value was obtained through a central-region analysis in a closed cubic box, which we call here “Big Box” (BB). In short if the simulation volume in SBC and PBC is V, we created a hard-walled simulation volume of side length $bbm \cdot V^{1/3}$ where bbm (“Big Box multiplier”) is an odd positive integer. Collisions would then be detected only in the central cubic region of volume V. As bbm increases the collision rate in the central region should approach the thermodynamic limit. Collisions would be detected by overlap and resolved by moving the colliders back to the position they occupied at the beginning of the timestep. This approach facilitates recursive resolution of collisions in crowded systems. For these specific simulations we used a volume fraction $\phi= 0.1$, N ranging from 2 to 2000 (for SBCs and PBCs) and bbm ranging from 3 to 29 (for BB). The simulations proceeded for a minimum of $10^{6}$ collisions (for SBCs and PBCs, $10^{5}$ for BB) and until convergence of the collision rate, as determined by a “two one sided t-tests” procedure with a tolerance of $10^{-6}$ over the last $10\%$ of the simulation time.\\

\textbf{Structural analysis}. To assess the effects of boundary shape and periodicity on the collective and local structure of the system we quantified a number of metrics. Random flight HSP simulations of N particles at various $\phi$ ($1\%, 2.5\%, 6.3\%, 15.9\%, 40\%$) were conducted for $10^{6}$ steps using both PBCs and SBCs. In addition, each condition was simulated for an interaction threshold $r_c=r$ and $r_c=0.2*R$, which would therefore determine the thickness of the “ghost halo” surrounding the simulation volume. For this set of simulations we used larger particles (r=10 nm, $\alpha = 0.7045$) to reduce the relative length of the typical displacement as compared to particle radius and therefore reduce the granularity of our simulations.
Every 10 timesteps the entire structure was probed: global pair distributions (in azimuth, elevation and distance) both with and without ghost particles; nearest-neighbor shell (NNS) pair distributions (in azimuth and elevation) were collected by thresholding the distance using the first trough of the radial pair distribution function (PDF) obtained from initial test runs; static structure factor (SSF) was calculated along the [100] family of directions, using wave vector values comprised between $2 \pi/L$ (where L is maximum size of the simulation volume along that direction) and $2 \pi/r$ in intervals of $2 \pi/L$. All data were stored for each snapshot for the purpose of determining the thermalization time.
The thermalization time was determined as follows. The azimuth distribution in the NNS was taken as diagnostic since the particles were positioned at the beginning of the simulations according to an fcc symmetry (analogous results were obtained from the corresponding elevation data). The moving average of the NNS azimuth distributions (window = 100 snapshots = 1000 timesteps, i.e., 100 azimuth bins were averaged separately over the window) was calculated leading to $10^{5}$ chronological distributions of NNS azimuths. The standard deviation (std) was calculated for each of them. Given the initial ordering of the particles the std started at a high value to then collapse into a steady distribution (cf. Figure S\ref{FigS1}A). The distribution of those standard deviations could be fitted accurately as a lognormal distribution (cf. Figure S\ref{FigS1}B) allowing an accurate determination of the confidence interval at $95\%$ of the mean. The thermalization was then determined as having occurred as soon as the std of the moving average of the azimuths in the NNS reached that confidence interval.\\

\textbf{Dynamic analysis}. To assess the effects of the boundary on the dynamics we conducted collisional simulations comparable to those used to determine collisional kinetics with an optimal N=1000 and $\phi=10^{-4}$, collecting a minimum of $10^{6}$ collisions whereby each collider was labeled. At the end of the simulations the intercollision waiting times distributions $\Psi$ were determined on a per-particle basis (i.e., $\Psi_{part}$, distribution of the times between any collision involving the i-th particle) and on a per-pair basis (i.e., $\Psi_{pair}$, distribution of the times between any collisions between the i-th and j-th particle).
To assess the different effects of PBCs and SBCs on the angular momentum we similarly analyzed the total angular momentum (L) for a system of 1000 Brownian particles.\\

\textbf{Benchmarking}. To compare the computational overhead of the SBC and PBC algorithms, we developed benchmark MATLAB scripts that isolate the core operations of each boundary condition. These non-parallelized codes performed a simplified loop consisting of: (i) displacing N particles with random vectors, (ii) applying the boundary condition logic, and (iii) calculating the all-pairs distance matrix. The cost of physical collision processing was intentionally excluded.
For the PBC benchmark, an optimized Minimum Image Convention (MIC) was used. The distance-wrapping correction was only applied to particle pairs where at least one particle was located within an interaction distance of the boundary; all other pair distances were computed using highly optimized native functions. For the SBC benchmark, given the absence of pre-existing optimized libraries, we developed and tested several implementations to ensure a robust comparison.
These benchmark tests were executed across multiple hardware architectures, and in all cases, they yielded consistent results for the relative performance between SBC and MIC-based PBC.

\section{Results and Discussion}

\subsection{COLLISIONAL KINETICS}

We first examined the collisional kinetics by measuring the collision rate as a function of particle number, N, for both SBC and PBC systems. As shown in Figure \ref{Fig1}B, both methods correctly converge to the same ground-truth value in the large-N limit, validating their baseline physical accuracy.
However, the paths to convergence are different: PBCs exhibits an apparently faster convergence rate. Our analysis reveals this is not a sign of higher efficiency, but is instead a consequence of spurious recurrence - a topological artifact where the periodic wrapping creates artificially short paths for particles to re-encounter one another. In contrast, SBC’s convergence rate reflects a more realistic kinetic process. This kinetic distinction, which will be further analyzed through waiting-time distributions, underscores the fundamental differences in how each boundary condition models system dynamics.

\subsection{STRUCTURAL ANALYSIS}
A structural analysis reveals profound differences in how SBC and PBC handle fluid organization, moving from subtle geometric effects to significant physical artifacts.\\

\begin{figure}
	\centering
	\includegraphics[width=0.5\linewidth]{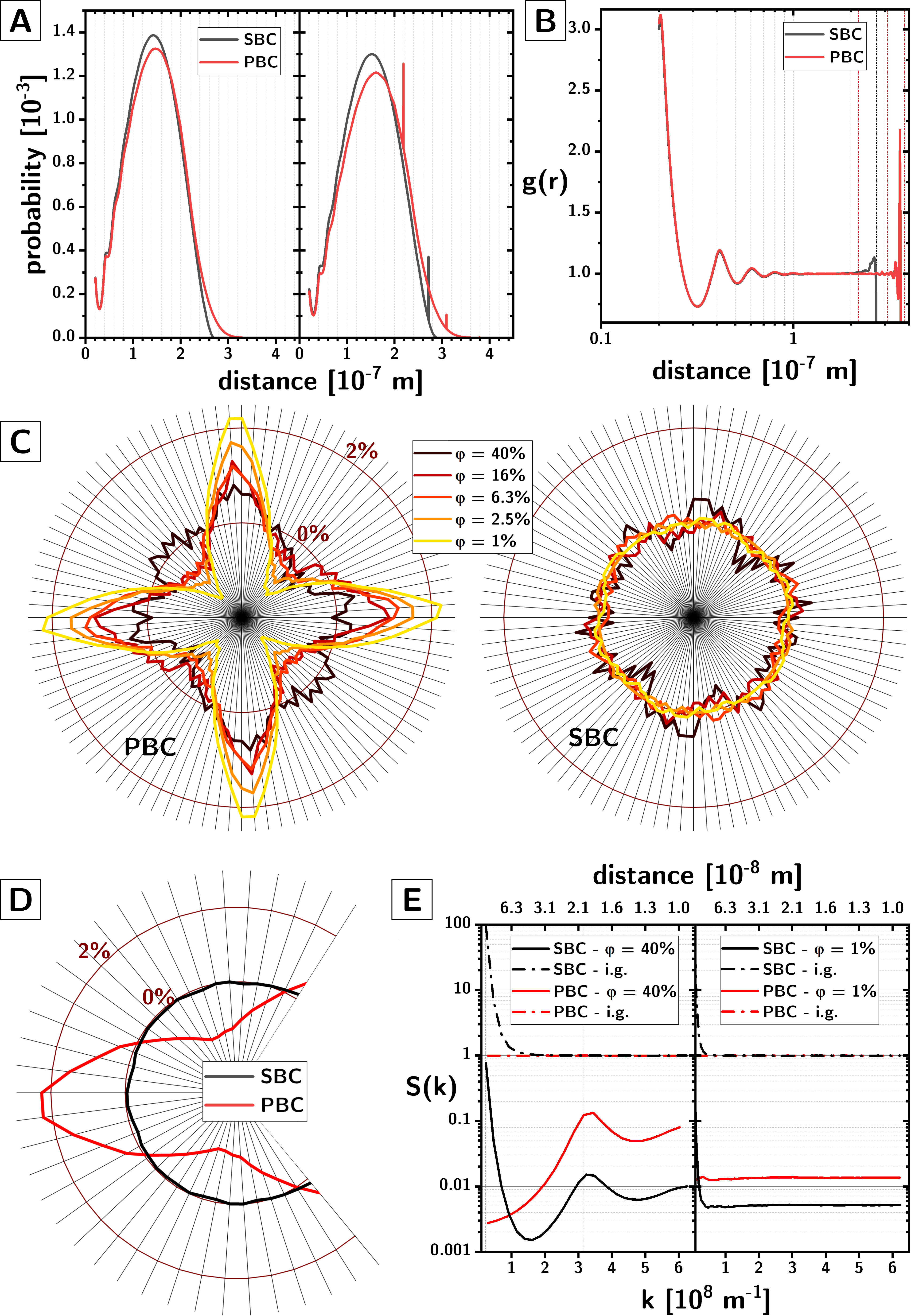}
	\caption{\textbf{Structural analysis}. \textbf{(A)} Raw pair-distance distributions (with, left, and without, right, ghost particles). \textbf{(B)} Pair distribution functions at $\phi=0.4$. \textbf{(C)}Radar plots of the deviation from isotropy of the particle/nearest-neighbor vector azimuth for PBC (left) and SBC (right) as a function of $\phi$. \textbf{(D)} Radar plots of the deviation from isotropy of the particle/nearest-neighbor vector elevation. \textbf{(E)} Static structure factor S(k) at $\phi=0.4$ (left) and $\phi=0.01$ (right). Dash-dotted lines show the expected ideal gas behaviour. In all panels, unless otherwise indicated, black lines indicate SBCs, red lines indicate PBCs.}
    \label{Fig2}
\end{figure}

\textbf{Distribution of pair distances}. The raw distribution of pair distances at $\phi=40\%$ (cf. Figure \ref{Fig2}A, left side without ghost particles, right side with ghost particles in HSP) shows expected differences. The SBC has a shorter and taller distribution since the diagonal of the cube is 1.396 times larger than the diameter of a sphere with the same volume. The distributions that include the ghost particle halo do show that the PBC conditions introduce the first spurious long-range correlation at a shorter distance than SBC.\\

\textbf{Pair-distribution functions}. The angle-averaged pair distribution functions (PDFs) show nearly identical short-range order for both methods (Fig. \ref{Fig2}B). This is a non-trivial finding, as one might expect the spherical boundary to induce sharper radial layering. However, distinct long-range artifacts emerge as the distance approaches the maximum separation distance. At high volume fractions ($\phi=40\%$), the SBC PDF exhibits a significant peak near 2R-2r. This peak, which has its own internal structure, is a many-body correlation caused by long-lived "traffic jams" of particles pinned by both a real particle near the boundary and its corresponding antipodal ghost.\\

The same many-body long-distance correlation occurs in the PBC PDF, but while the SBC PDF would monotonically increase in proximity of 2R-2r, the PBC PDF oscillates wildly above and below 1 before the divergence at $L\sqrt3$. This is a strong indication of an ordered long range correlation between the two collisional “traffic jams” mentioned above, which is only explainable with a shared preferred orientation and symmetry of the particle clusters.\\

\textbf{Nearest-neighbor-shell anisotropy}. This orientational correlation is the most profound structural difference between PBC and SBC. It was revealed by the distributions of the azimuths of the vectors connecting any particle to its nearest neighbors as a function of $\phi$. The radar plots in Figure \ref{Fig2}C show the deviation from the ideal isotropic case of the two boundary conditions. The SBC system remains perfectly isotropic even at the highest volume fraction of $40\%$ (cf. Figure \ref{Fig2}C, right). In stark contrast, the PBC system exhibits significant anisotropy even at a low volume fraction of $1\%$ (cf. Figure \ref{Fig2}C, left). The distributions show a clear four-fold symmetry, with particle packing preferentially aligned along the artificial axes of the PBC cubic lattice. This provides definitive evidence that PBCs imposes a non-physical orientational order on the fluid, an artifact that SBCs, by their spherical nature, completely eliminate.\\

The same information is shown in the elevation data (cf. Figure \ref{Fig2}D) for which PBCs yield an increased probability for 0 elevations, while SBCs yield a distribution comparable to the ideal gas analogue.\\

\textbf{Static structure factors}. Lastly, we characterized the static structure factor S(k) along the [100] direction for both boundary conditions. The plots in Figure \ref{Fig2}E include the corresponding plots obtained from an ideal gas simulation (dash-dotted lines). Qualitatively the S(k) from the SBC and PBC simulations at $\phi= 40\%$ are similar (cf. Figure \ref{Fig2}E, left): the high k behaviour show self-correlation and a peak corresponding to the NNS (dash-dotted vertical line). The SBC is distinct for two traits: (i) the divergence at low k matching that of the ideal gas analogue and therefore due to the finite spherical boundary, and (ii) the overall significantly lower value of S(k) as compared to PBCs.
It is expected for crowded systems to manifest their large excluded volumes with a broadband suppression of fluctuations which is then picked up by the SSF. In fact, the S(k) of PBCs is also 1-2 orders of magnitude lower than the ideal gas analogue. The even lower value of S(k) for the SBC simulations is, in our attribution, the result of the increased homogeneity demonstrated in Figure \ref{Fig2}C-D. The cubic structuring caused by PBCs increases the average fluctuation density in the system. This same last effect is evident even when $\phi$ is decreased to $1\%$ (cf. Figure \ref{Fig2}E, right) even though, consistently with our attribution above, is much less evident ($S_{PBC(k)/SSBC(k)} \sim 9$ at $\phi=40\%$ while $S_{PBC(k)}/S_{SBC(k)} \sim 2.5$ at $\phi=1\%$).

\subsection{DYNAMIC ANALYSIS}

\begin{figure}
	\centering
	\includegraphics[width=0.55\linewidth]{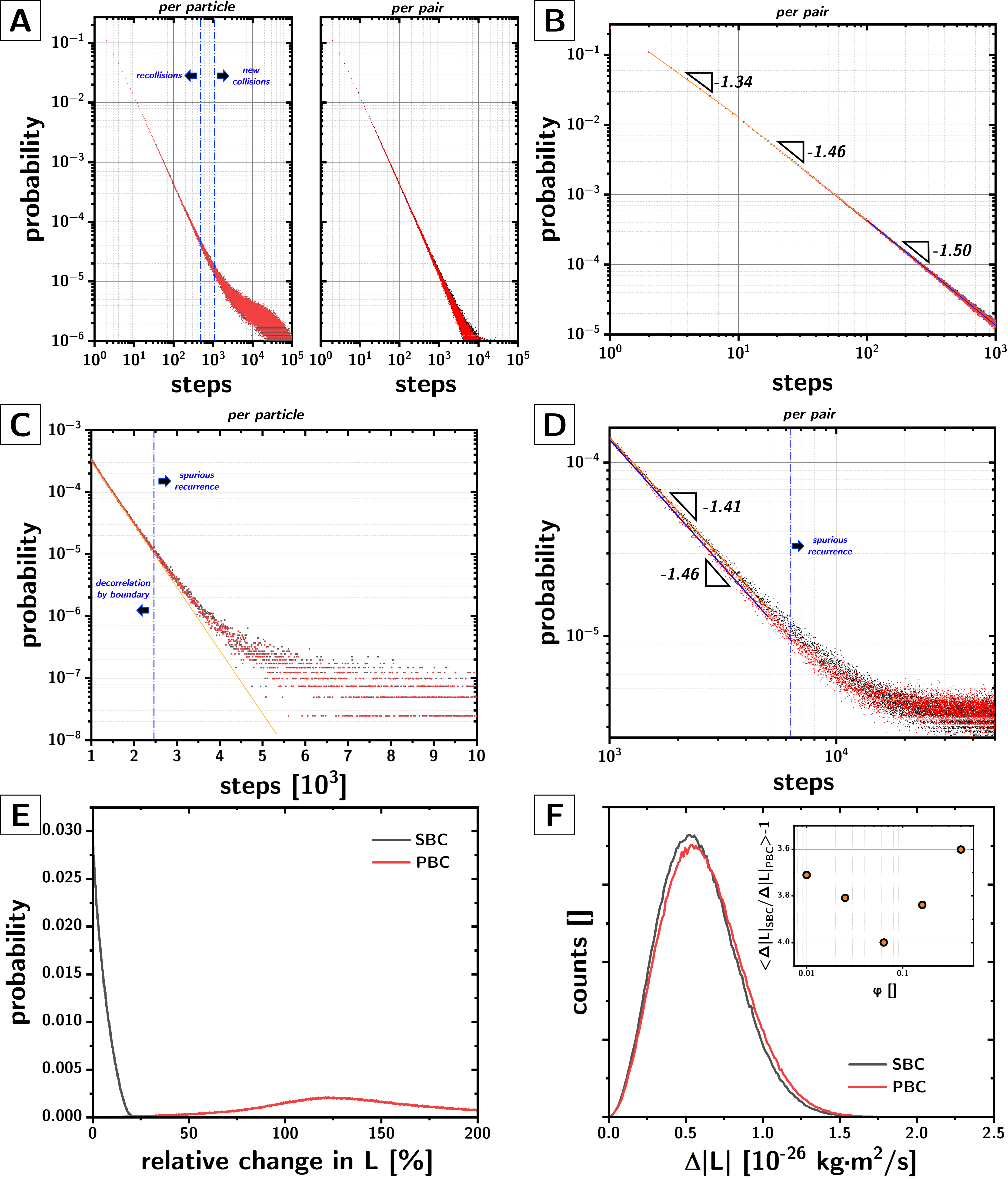}
	\caption[erfdv]{\textbf{Dynamics analysis}. \textbf{(A)} Loglog plots of the per-particle (left) and per-pair (right) intercollision time distributions for SBCs (black) and PBCs (red) simulations ($\phi=10^{-4}, N=1000)$. \textbf{(B)} Zoomed-in loglog plot of the 1-103 timesteps range of the per-pair intercollision time distributions highlighting the evolving exponent of the power-law dependence. \textbf{(C)} Zoomed-in semilog plot of the range >$10^{3}$ timesteps of the per-particle intercollision time distributions highlighting the exponential Smoluchowski range and the spurious recurrence regime above $\sim 2500$ timesteps. \textbf{(D)} Zoomed-in loglog plot of the range $>10^{3}$ timesteps of the per-pair intercollision time distributions highlighting the deviation above power-law due to recurrence. \textbf{(E)} Relative magnitude change in angular momentum associated with a boundary crossing. \textbf{(F)} Distribution of global angular momentum changes on a timestep basis. In the insect the relative difference (in $\%$) of the average changes in global angular momentum between PBC and SBC.}
    \label{Fig3}
\end{figure}

Intercollision waiting times distributions. The distributions $\Psi_{part}$ and $\Psi_{pair}$ of the intercollision times on a per-particle (cf. Figure \ref{Fig3}A, left) and per-pair (cf. Figure \ref{Fig3}A, right) basis (N=1000, $\phi=10^{-4}$) provide a wealth of information about how the long-term dynamics of the particles are affected by the boundary conditions. Figure \ref{Fig1}A shows clearly that, at short times, both $\Psi_{part}$ and $\Psi_{pair}$ are power-law distributed as expected by first passage statistics\cite{55}. As time reaches a certain threshold, the $\Psi_{part}$ rises above the power law. This crossover marks the transition from a dynamic regime dominated by re-collisions to one governed by random, memoryless encounters with new partners, which follows exponential statistics. The fact that the $\Psi_{pair}$ shows no such deviation confirms this interpretation (cf. Figure \ref{Fig1}A, right). 
Furthermore, we observe a deviation from ideal point-particle theory at very short timescales (Figure \ref{Fig3}B). The initial power-law exponent is less negative than the theoretical -3/2 value, indicating that the finite size of the particles enhances the probability of immediate re-collisions. As the time delay increases, the particles begin to behave more like points relative to their separation distance, and the exponent correctly relaxes towards the expected -3/2.
Critically, in this short-delay regime governed by local physics, we find no significant difference between the $\Psi$ distributions for SBC and PBC. This is expected, as the influence of the boundary's global shape and topology should only manifest at longer timescales, when diffusing particles have had sufficient time to probe the full extent of the simulation volume.
As mentioned above, the $\Psi_{part}$ deviates from the power-law-distributed recollisions when new colliders enter the fray, recovering the Smoluchowski statistics. This is shown in Figure \ref{Fig3}C, where the semilog plot of $\Psi_{part}$ starting from 1000 steps is clearly linear and largely independent of boundary conditions. On the other hand, past 2500 steps we observed a further regime establish by a deviation above the exponential trend. This last regime is caused by spurious recurrence and therefore by the direct intervention of the boundary rules on the statistics of encounters. Essentially, particles are prevented from being further apart than 2R or $L\sqrt3$, which inevitably causes a spuriously high probability of recollisions of the same particles at very long times.
We isolated this behaviour by looking instead at the $\Psi_{pair}$. In an infinite system the $\Psi_{pair}$ would remain a power-law at any time interval. The spurious recurrence caused by the boundaries instead causes a deviation when the average distance between a collided pair that has separated crosses the size of the boundary. To achieve sufficient statistics to observe this very long time-scale behaviour in $\Psi_{pair}$ we ran a targeted simulation at a higher $\phi=1\%$ and looking at $10^{7}$ total collisions. The data (cf. Figure \ref{Fig3}D) shows two things: (i) SBC starts deviating from the power law slightly before PBC, but (ii) PBC remains higher than SBC after the deviation. The first observation can be understood, in our opinion, as a byproduct of the curvature of the boundary wall. The concave survature of the SBCs slightly increases (depending on $\phi$) the probability that a collided pair would get decorrelated by a boundary crossing. The second observation can be understood, in our opinion, with the increased average recurrence in PBCs caused by the fact that on average, ghost particles are closer to their real counterparts in PBCs than in SBCs.
Conservation of angular momentum. Finally, we analyzed the conservation of total angular momentum, L, a key indicator of a simulation's physical fidelity. We first quantified the relative change $\Delta L/L$ caused by a single particle crossing the boundary. The distributions of these changes for SBCs and PBCs reveal a stark contrast (Figure \ref{Fig3}E). The coordinate-wrapping algorithm in PBC induces a significant, unphysical impulse with each crossing. In contrast, the handover protocol in SBC, which is designed for kinematic continuity, results in a $\Delta L$ that is much smaller and centered at zero.
This microscopic difference in the boundary algorithms has a direct macroscopic consequence on the global dynamics. While BD simulations are inherently non-conservative due to the thermostat, we can isolate the additional error from the boundary condition by analyzing the equilibrium fluctuations of the total angular momentum (Figure \ref{Fig3}F). Our data show that the PBC simulations exhibits a systematic $\sim 4\%$ larger average $\Delta L / L$ as compared to the SBC simulation, regardless of volume fraction. This demonstrates that the PBC wrapping algorithm continuously injects a non-insignificant amount of unphysical, algorithmic noise into the system’s rotational dynamics—an artifact that SBC eliminates. 

\section{BENCHMARKING}

\begin{figure}
	\centering
	\includegraphics[width=0.3\linewidth]{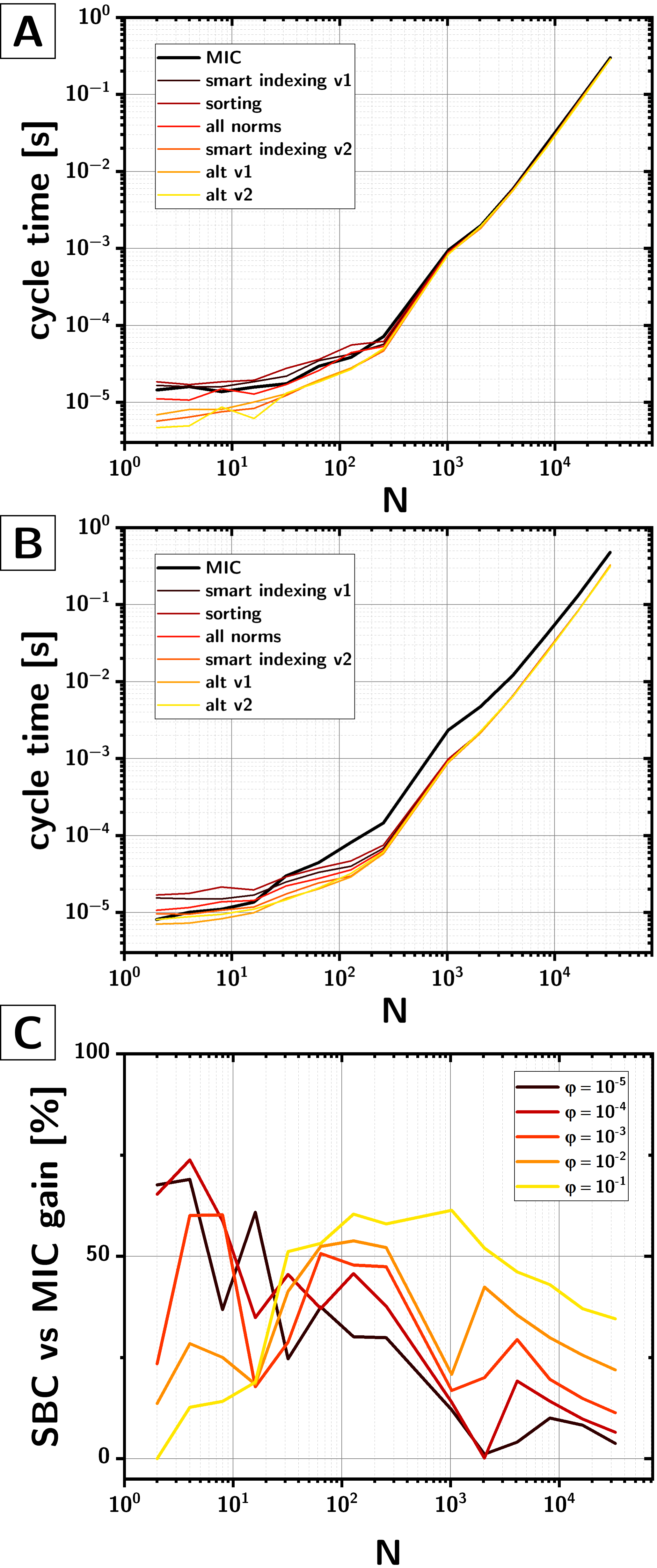}
	\caption{\textbf{Benchmarking}. \textbf{(A)} Execution times of a single timestep for Brownian dynamics simulations at $\phi=10^{-5}$ employing MIC-based PBCs and a panel of SBCs algorithms as a function of N. \textbf{(B)} Same as panel A but using $\phi=10^{-1}$. \textbf{(C)} $\%$ gain in execution speed of SBC as compared to MIC as a function of N and $\phi$. }
    \label{Fig4}
\end{figure}

Finally, we benchmarked the computational performance of our SBC implementation against a highly optimized Minimum Image Convention (MIC) algorithm, which represents the computational state-of-the-art for PBC. The results, shown in Figure \ref{Fig4}, reveal a strong dependence on system density.
At very low volume fractions ($\phi= 10^{-5}$), the performance of SBC and MIC are nearly indistinguishable, as boundary interactions are rare events (Fig. \ref{Fig4}A). However, as the system becomes more crowded, SBCs become significantly more performant than MIC. At a volume fraction of $\phi = 0.1$, SBC demonstrates a clear advantage across a wide range of particle numbers, N (Fig. \ref{Fig4}B).
The largest performance gains, with SBC being up to $\sim 60 \%$ faster than the MIC-based PBC, were observed for crowded systems ($ \phi> 10^{-3}$) with intermediate particle counts (N between $10^{3}$ and $10^{4}$). This observation suggests that for the crowded systems where boundary condition artifacts are most problematic, SBC is not only more physically accurate but also computationally more efficient. 

\section{Conclusion}

In this work, we introduced and validated Spherical Boundary Conditions (SBC), a method for particle simulations designed as a high-fidelity alternative to traditional Periodic Boundary Conditions (PBC) for isotropic systems. Our comparison using hard-sphere Brownian Dynamics reveals that the distinct geometries and boundary rules of SBC and PBC result in systems with fundamentally different physical properties and computational performance, particularly in the crowded regimes central to chemical physics.
We have demonstrated that the primary advantage of SBC is its perfect isotropy. Unlike PBC, whose cubic lattice imposes a non-physical anisotropic order on the fluid structure and dynamics, SBC produces a system free of these directional artifacts. This was confirmed by an angle-resolved analysis of the nearest-neighbor shell and the Static Structure Factor, S(k). Furthermore, we have shown that SBC provides a superior conservation of total angular momentum, eliminating the systematic, algorithmic noise inherent to the coordinate-wrapping protocol of PBC.
Crucially, for the target regime of crowded systems ($\phi> 1\%$), where the artifacts of PBC are most severe, we found SBC to be not only more physically accurate but also computationally more efficient, outperforming standard MIC-based implementations by up to $60\%$.
Therefore, we conclude that SBC is a powerful and efficient new tool for the community. For studies where isotropy, accurate structural determination, and correct rotational dynamics are paramount, SBC might presents a superior alternative to traditional periodic methods for simulating crowded, disordered matter. Work is ongoing to explore the implementation of soft potentials and anisotropic shapes in this framework.

\section{Acknowledgements}

This research benefits from the High Performance Computing facility of the University of Parma, Italy (\url{https://www.hpc.unipr.it}).

\newpage
 
\pagestyle{fancy}
\fancyhf{}
\fancyhead[LO]{SUPPORTING INFORMATION: MODELING BROWNIAN MOTION AS A TIMELAPSE OF THE PHYSICAL TRAJECTORY}
\fancyfoot[C]{\thepage}

\renewcommand{\thefigure}{S\arabic{figure}}
\setcounter{figure}{0}

\begin{titlepage}
\centering
{\LARGE\bfseries SUPPORTING INFORMATION: PSEUDOPERIODIC SPHERICAL BOUNDARY CONDITIONS: EFFICIENT AND ISOTROPIC 3D PARTICLE SIMULATIONS WITHOUT LATTICE ARTIFACTS \par}

\vspace{0.5cm}

{\large
\textbf{MANUEL DEDOLA$^{\dag}$, LUDOVICO CADEMARTIRI$^{\dag *}$} \\
$^{\dag}$Department of Chemistry, Life Sciences and Environmental Sustainability, University of Parma,\\
Parco Area delle Scienze 17 A, Parma, Italy. \\
$^{*}$Author to whom correspondence should be addressed: \\
ludovico.cademartiri@unipr.it \par
}

\end{titlepage}

\RaggedRight

\setcounter{section}{0}
\section{Supporting Information}

\textbf{Codes}. All simulations were conducted with homebrew code in Matlab running on a range of architectures. Most results were validated by two separate codes independently written by the two authors.

\textbf{Physical parameters}. The simulations were conducted considering silica particles in water using the following parameters.

\begin{center}
\rowcolors{2}{gray!25}{white}
\begin{tabular}{|l|l|l|}
\hline
\textbf{Property} & \textbf{Value} &\textbf{ Units} \\
\hline
\textit{Particle density} & 2285 & $\mathrm{Kg} \cdot \mathrm{m}^{-3}$ \\
\hline
\textit{Solvent mass} & $2.9916 \cdot 10^{-26}$ & Kg \\
\hline
\textit{Solvent density} & 1000 & $\mathrm{Kg} \cdot \mathrm{m}^{-3}$ \\
\hline
\textit{Solvent dynamic viscosity} & $8.9 \cdot 10^{-4}$ & $Pa \cdot s$ \\
\hline
\textit{Solvent vorticity timescale} & $1.4318 \cdot 10^{-12}$ & s \\
\hline
\textit{Solvent speed of sound} & 1498 & $\mathrm{m} \cdot \mathrm{s}^{-1}$ \\
\hline
\textit{Temperature} & 298.15 & K \\
\hline
\end{tabular}
\end{center}

\textbf{Timescales and lengthscales}. Brownian motion was described as a Markovian random flight. The choice of timescales and step lengths was based on our recent work that details how to quantitatively represent a physical Brownian trajectory as a random walk.\cite{1s} In short, a physical trajectory is generated on timescales commensurate with the collisional timescale $\tau_{c}$ (i.e., the average interval in time between consecutive collisions of a particle with any solvent molecules at its surface) estimated as

\begin{equation*}
\tau_{c} \cong \frac{9.775 \cdot 10^{-11}}{\sqrt{k_{B} T}}\left(\frac{r_{s}}{r_{p}}\right)^{2} \frac{\sqrt{m_{s} m_{p}}}{\sqrt{m_{s}}+\sqrt{m_{p}}} \tag{1}
\end{equation*}

The trajectory is created by autoregressing the velocity autocorrelation function (VACF) that accounts for hydrodynamic and acoustic effects leading to the well-known algebraic tail with an exponent equal to $-3 / 2$ \cite{2s,3s,4s,5s,6s,7s,8s,9s}.

\begin{equation*}
C_{v}(t)=\frac{k_{B} T}{M}\left\{\frac{2 \rho_{p}}{9 \pi \rho_{s}} \int_{0}^{\infty} \frac{\sqrt{x} e^{-x t / \tau_{v}}}{1+\sigma_{1}+\sigma_{2} x^{2}} d x+\frac{e^{\alpha_{1} t / \tau_{s}}}{1+2 \rho_{p} / \rho_{s}}\left[\cos \left(\frac{\alpha_{2} t}{\tau_{s}}\right)-\frac{\alpha_{1}}{\alpha_{2}} \sin \left(\frac{\alpha_{2} t}{\tau_{s}}\right)\right]\right\} \tag{2}
\end{equation*}

where $\rho_{\mathrm{p}}$ is the density of the particle, $\rho_{\mathrm{s}}$ is the density of the solvent, $\sigma_{1}=\frac{1}{9}\left(7-4 \rho_{\mathrm{p}} / \rho_{\mathrm{s}}\right)$, $\sigma_{2}=\frac{1}{81}\left(1+2 \rho_{p} / \rho_{s}\right)^{2}, \alpha_{1}=1+\rho_{s} / 2 \rho_{p}, \alpha_{2}=\sqrt{1-\left(\rho_{s} / 2 \rho_{p}\right)^{2}}$, and $M$ is the "effective mass" of the particle $M=\frac{4}{3} \pi r_{p}^{3}\left(\rho_{p}+\frac{\rho_{s}}{2}\right)$.

From the calculated VACF we could calculate the MSD corresponding to time $\tau_{c}$ which was used as the variance of the single step. The autocorrelation was introduced by back calculating from the VACF the autoregression coefficients\cite{10s} $\theta_{\lambda}$ for the expansion

\begin{equation*}
\mathrm{v}_{\mathrm{x}}(\mathrm{t})=\sum_{\lambda=1}^{\lambda_{\max }} \beta_{\lambda} \mathrm{v}(\mathrm{t}-\lambda)+\xi(\mathrm{t}) \tag{3}
\end{equation*}

where $v_{x}(t)$ is the velocity time series (and hence the single timestep displacement is $\left.d x(t)=v_{x}(t) \cdot \tau_{c}\right), \lambda$ is the lag, and $\xi(t)$ is a gaussian noise function. The autoregression is performed by using the DurbinLevinson algorithm\cite{11s}.

Once we determined the autoregression coefficients, any number of physical trajectories can be calculated ex novo knowing that they will obey the VACF above.

As shown in our prior work, this approach allows to determine the timescale for full decorrelation of the physical trajectory in terms of the relaxation time $\tau_{r}$. The relaxation time is generally estimated as

\begin{equation*}
\tau_{r}=\frac{2}{9} \frac{\rho_{p}}{\eta} r_{p}^{2} \tag{4}
\end{equation*}

for spheres\cite{12s} (i.e., Einstein-Stokes approximation) , where $\eta$ is the dynamic viscosity of the solvent and $r_{p}$ is the radius of the particle. In our case, given that the decay in momentum is not entirely given by solvent-particle collisions, we performed a Green-Kubo integration of the VACF

\begin{equation*}
\tau_{r}=\frac{1}{C_{v}(0)} \int_{0}^{\infty} C_{v}(t) d t \tag{5}
\end{equation*}

Our findings indicate that the physical trajectory can be considered mostly decorrelated only after ${ }^{\sim} 200 \tau_{\mathrm{r} .}$ \cite{1s}

This result was obtained by taking timelapses of the physical trajectory every $200 \tau_{r}$, and then looking at the autocorrelation of the subsampled trajectory. A crucial observation was that the diffusivity extracted from such subsampled trajectories (simply by dividing the MSD in one coordinate by two times the timestep) was significantly smaller than the Einstein-Stokes diffusivity.

Consequently we discovered that when describing physical Brownian trajectories as random flights one has to correct the Einstein-Stokes diffusivity $\mathrm{D}_{\mathrm{E}-\mathrm{S}}$ or risk overestimating the overall displacement of the particles over time.

The challenge, as described in the paper, is that there is no analytical expression for the correction to the diffusivity ( $\alpha=D_{\text {eff }} / D_{E-S}$ ) and therefore it must be numerically calculated for each condition.

In this paper we considered mostly silica particles of 1.12 nm radius in water for which the $\alpha$ was calculated as 0.7074 when the timestep with the random walk was taken to be $200 \tau_{\mathrm{r}}$. For the structural analysis we used larger particles of 10 nm radius in water for which the $\alpha$ was calculated as 0.7045 .

The determination of the $\alpha$ parameter for larger particles requires one to generate the physical trajectories using a coarser timescale. The problem is a computational one. As the particles increase in size, the collisional timescale $\tau_{c}$ decreases while $\tau_{r}$ increases. Overall, the number of $\tau_{c}$ timesteps one needs to generate to model a single $\tau_{r}$ interval scales with $\sim r^{4}$. So, for a 10 nm particle you could easily need $10^{6} \tau_{c}$ steps to calculate a single $\tau_{r}$. Since the VACF is very long tailed, we take subsamples every $200 \tau_{r}$, which means that in order to calculate a single random flight displacement we would need to calculate in excess of $10^{8}$ steps, all autocorrelated with each other. So, while the $\alpha$ for the 1.12 nm particles was calculated with a genuine collisional timescale (Eq. 1), the $\alpha$ for the 10 nm particles was calculated by using as $\tau_{c}$ a fixed portion of $\tau_{r}$ (specifically $1000 \tau_{c}=\tau_{r}$ ) with the physical meaning that this collisional timescale does not represents a single collision with the solvent molecule but a fixed degree of decorrelation.

Collision handling. The collisions were handled by resetting the particle to the position they had at the beginning of the step.

\section*{MATERIALS}
Four architectures were used over the course of this study

\section*{Workstation 1}

\subsection{System}

\rowcolors{2}{gray!25}{white}
\begin{tabular}{|l|l|}
\hline Item & Value \\
\hline OS Name & Microsoft Windows 11 Pro \\
\hline System Manufacturer & Gigabyte Technology Co., Ltd. \\
\hline System Model & X670 GAMING X AX \\
\hline System Type & x64-based PC \\
\hline Processor & AMD Ryzen 9 7950X, 16C/32T @ 4.5 GHz \\
\hline BIOS Version/Date & American Megatrends International, LLC. F21, 1/10/2024 \\
\hline SMBIOS Version & 3.6 \\
\hline Embedded Controller Version & 255.255 \\
\hline BIOS Mode & UEFI \\
\hline BaseBoard Manufacturer & Gigabyte Technology Co., Ltd. \\
\hline BaseBoard Product & X670 GAMING X AX \\
\hline BaseBoard Version & x.x \\
\hline Installed Physical Memory (RAM) & 96.0 GB \\
\hline Total Physical Memory & 95.1 GB \\
\hline Available Physical Memory & 63.3 GB \\
\hline Total Virtual Memory & 258 GB \\
\hline Available Virtual Memory & 213 GB \\
\hline Page File Space & 163 GB \\
\hline
\end{tabular}

\subsection{Display}

\begin{tabular}{|l|l|}
\hline Item & Value \\
\hline Name & AMD Radeon (TM) Graphics \\
\hline Adapter Type & AMD Radeon Graphics Processor ( $0 \times 164 \mathrm{E}$ ), Advanced Micro Devices, Inc. compatible \\
\hline Adapter Description & AMD Radeon (TM) Graphics \\
\hline Adapter RAM & $512.00 \mathrm{MB}(536,870,912$ bytes) \\
\hline Driver Version & 32.0.11024.2 \\
\hline Name & AMD Radeon PRO W6600 \\
\hline
\end{tabular}

\subsection{Disks}

\begin{center}
\begin{tabular}{|l|l|}
\hline
Item & Value \\
\hline
Description & Disk drive \\
\hline
Manufacturer & (Standard disk drives) \\
\hline
Model & ST18000NM000J-2TV103 \\
\hline
Bytes/Sector & 512 \\
\hline
Media Loaded & Yes \\
\hline
Media Type & Fixed hard disk \\
\hline
Partitions & 1 \\
\hline
Size & $16.37 \mathrm{~TB}(18,000,202,752,000$ bytes) \\
\hline
Description & Disk drive \\
\hline
Manufacturer & (Standard disk drives) \\
\hline
Model & Samsung SSD 990 PRO 2TB \\
\hline
Media Type & Fixed hard disk \\
\hline
Partitions & 3 \\
\hline
Size & 1.82 TB (2,000,396,321,280 bytes) \\
\hline
\end{tabular}
\end{center}

\section*{Workstation 2}

\begin{center}
\begin{tabular}{|l|l|}
\hline
\textbf{Item} & \textbf{Value} \\
\hline
OS Name & Linux \\
\hline
JOB management & SLURM \\
\hline
System Manufacturer & Scientific Computing of the University of Parma (HPC) \\
\hline
Partition & cpu \\
\hline
Node name & wn18-wn19 \\
\hline
Processor & 4 Intel Xeon E5-6140 2.3GHz 19c \\
\hline
\#Cores available & 72 \\
\hline
MFM (GB) available & 384 \\
\hline
\# Total cores & 144 \\
\hline
\# of nodes used & 1 \\
\hline
MEM (GB) used & 125 \\
\hline
Job time (hr) & 72 \\
\hline
\end{tabular}
\end{center}

\section*{Workstation 3}

\begin{center}
\begin{tabular}{|l|l|}
\hline
\textbf{Item} & \textbf{Value} \\
\hline
OS Name & Linux \\
\hline
JOB management & SLURM \\
\hline
System Manufacturer & Scientific Computing of the University of Parma (HPC) \\
\hline
Partition & cpu \\
\hline
Node name & Wn 34 \\
\hline
Processor & 4 Intel Xeon E7-8880v4 2.2 GHz 22 C \\
\hline
\#Cores available & 88 \\
\hline
MFM (GB) available & 1024 \\
\hline
\# Total cores & 88 \\
\hline
\# of nodes used & 1 \\
\hline
MEM (GB) used & 125 \\
\hline
Job time (hr) & 72 \\
\hline
\end{tabular}
\end{center}

\section*{Workstation 4}

\begin{center}
\begin{tabular}{|l|l|}
\hline
\textbf{Item} & \textbf{Value} \\
\hline
OS Name & Linux \\
\hline
JOB management & SLURM \\
\hline
System Manufacturer & Scientific Computing of the University of Parma (HPC) \\
\hline
Partition & cpu \\
\hline
Node name & Wn 35-36 \\
\hline
Processor & 4 Intel Xeon E5-6252n 2.3GHz 24c \\
\hline
\#Cores available & 96 \\
\hline
MEM (GB) available & 1024 \\
\hline
\# Total cores & 192 \\
\hline
\# of nodes used & 1 \\
\hline
MEM (GB) used & 125 \\
\hline
Job time (hr) & 72 \\
\hline
\end{tabular}
\end{center}

\begin{figure}
	\centering
	\includegraphics[width=0.7\linewidth]{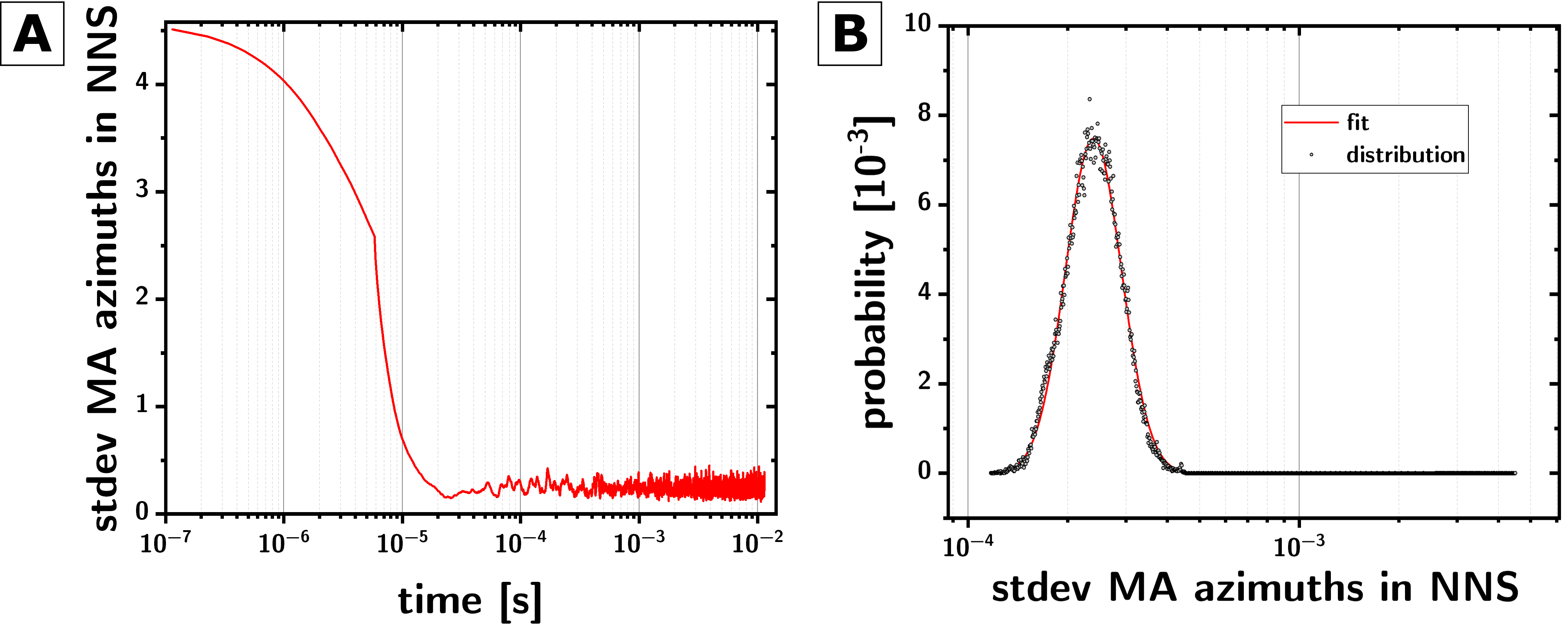}
	\caption{\textbf{Thermalization}. (\textbf{A}) Standard deviation of the moving average of the azimuths of the particle/nearest-neighbour-shell vectors as a function of time. \textbf{(B)} Distribution of the data in panel (A) showing clear lognormal distribution. The 95\% of the mean value of the distribution were used as thresholds to signal thermalization.}
    \label{FigS1}
\end{figure}

\section*{DERIVATIONS}
\subsubsection{\textit{Derivation \#1}. Dependence between the displacements of real and ghost particles in SBC}

\begin{figure}
	\centering
	\includegraphics[width=0.45\linewidth]{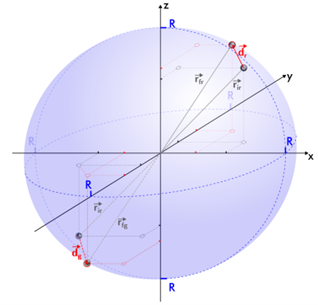}
\end{figure}

Assuming the following known variables: the radius of the simulation volume $\mathbf{R}$, the position of the real particle $\boldsymbol{r}_{i, r}$ and its displacement $\boldsymbol{d}_{\boldsymbol{r}}$ we can calculate the position of the ghost particle as well as its displacement.

Final position of real particle:
\begin{equation}
    \overrightarrow{r_{f, r}}=\overrightarrow{r_{l, r}}+\overrightarrow{d_{r}}
\end{equation}
Initial position of the ghost particle:
\begin{equation}
\overrightarrow{r_{l, g}}=\overrightarrow{r_{l, r}}-2 R \frac{\overrightarrow{r_{l, r}}}{\left|\overrightarrow{r_{l, r}}\right|}
\end{equation}

Final position of the ghost particle:

\begin{equation}
\overrightarrow{r_{f, g}}=\overrightarrow{r_{f, r}}-2 R \frac{\overrightarrow{r_{f, r}}}{\left|\overrightarrow{r_{f, r}}\right|}
\end{equation}

Displacement of ghost particle:\\

$$
\begin{gathered}
\overrightarrow{d_{g}}=\overrightarrow{r_{f, g}}-\overrightarrow{r_{l, g}}=\left(\overrightarrow{r_{f, r}}-2 R \frac{\overrightarrow{r_{f, r}}}{\left|\overrightarrow{r_{f, r}}\right|}\right)-\left(\overrightarrow{r_{l, r}}-2 R \frac{\overrightarrow{r_{l, r}}}{\left|\overrightarrow{r_{l, r}}\right|}\right)= \\
=\left(\overrightarrow{r_{l, r}}+\overrightarrow{d_{r}}-2 R \frac{\overrightarrow{r_{l, r}}+\overrightarrow{d_{r}}}{\left|\overrightarrow{r_{l, r}}+\overrightarrow{d_{r}}\right|}\right)-\left(\overrightarrow{r_{l, r}}-2 R \frac{\overrightarrow{r_{l, r}}}{\left|\overrightarrow{r_{l, r}}\right|}\right) \\
\overrightarrow{d_{g}}=\overrightarrow{d_{r}}+2 R\left(\frac{\overrightarrow{r_{l, r}}}{\left|\overrightarrow{r_{l, r}}\right|}-\frac{\overrightarrow{r_{l, r}}+\overrightarrow{d_{r}}}{\left|\overrightarrow{r_{l, r}}+\overrightarrow{d_{r}}\right|}\right)
\end{gathered}
$$

Importantly, the angle $\xi$ between the displacement of the real particle and that of the ghost particle can be easily derived as\\

$$
\xi=\arccos \left(\frac{\left|\overrightarrow{d_{r}}\right|^{2}+2 R\left(\overrightarrow{d_{r}} \cdot \widehat{r_{l, r}}-\overrightarrow{d_{r}} \cdot \widehat{r_{f, r}}\right)}{\left|\overrightarrow{d_{r}}\right| \sqrt{\left|\overrightarrow{d_{r}}\right|^{2}+4 R\left(\overrightarrow{d_{r}} \cdot \widehat{r_{l, r}}-\overrightarrow{d_{r}} \cdot \widehat{r_{f, r}}\right)+8 R^{2}\left(1-\widehat{r_{l, r}} \cdot \widehat{r_{f, r}}\right)}}\right)
$$

If the displacement of the real particle is radial, then the angle $\xi$ is zero. The angle reaches a maximum $(\pi)$ for displacement that is orbital with respect to the origin, as shown in Figure below (where "displacement elevation" is the angle (from $-\pi / 2$ to $\pi / 2$ ) between the displacement vector and the plane tangent to the boundary at the point of intersection between the direction of $\widehat{r_{l, r}}$, so equal to zero if the displacement is orbital and equal to $\pm \pi / 2$ if the displacement is radial).\\

\begin{figure}
	\centering
	\includegraphics[width=0.5\linewidth]{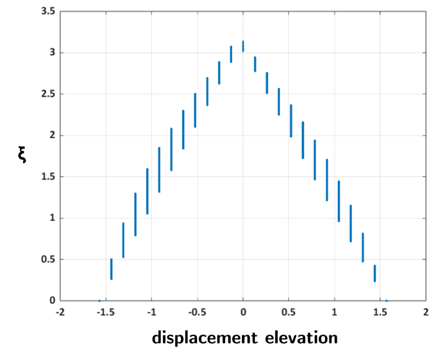}
\end{figure}

\subsection{\textit{Derivation \#2}. Estimation of volume loss in SBC\\}

\begin{figure}[H]
	\centering
	\includegraphics[width=0.5\linewidth]{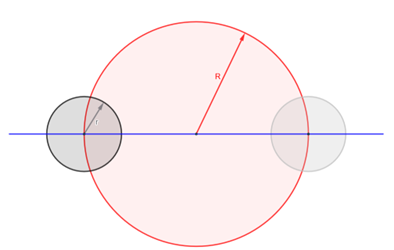}
\end{figure}

Given that the particle size is much smaller than the radius of the spherical box, the local surface can be considered approximately flat. In the worst-case scenario - where the center of mass of the particle lies exactly on the boundary of the spherical box - we estimate the volume loss fraction $f$ as:

$$
f(r, R)=\frac{V}{V_{p}}
$$

Here, $V=\frac{2 \pi}{3}\left[\left(\frac{r^{2}}{2 R}\right)^{2}\left(3 R-\frac{r^{2}}{2 R}\right)+\left(r-\frac{r^{2}}{2 R}\right)^{2}\left(2 r+\frac{r^{2}}{2 R}\right)\right]$ is the volume of the particle enclosed within the box, and $V_{p}=\frac{4}{3} \pi r^{3}$ is the total volume of a spherical particle with radius r .

The value of $f$ ranges from 0 to 1 , where values closer to 1 indicate minimal volume loss and better confinement of the particle within the spherical box. The specific $f$ value used in our simulations are reported in Table T1.

TABLE T1. Some values of volume loss fraction used in the simulations. As expected, an increase in $\phi$ and in the number of particles leads to an increase in the size of the box and consequently to an increase in volume loss fraction.

\begin{center}
\begin{tabular}{|l|l|l|l|}
\hline
$\phi$ & $N_{p}$ & $\boldsymbol{r} \boldsymbol{[} \boldsymbol{m} \boldsymbol{]}$ & $f$ \\
\hline
$10^{-1}$ & 212 & $1.00 \cdot 10^{-8}$ & 0.971 \\
\hline
$10^{-3}$ & 45 & $5 \cdot 10^{-10}$ & 0.990 \\
\hline
$10^{-2}$ & 2 & $5 \cdot 10^{-10}$ & 0.994 \\
\hline
$10^{-1}$ & 212 & $2.24 \cdot 10^{-9}$ & 0.971 \\
\hline
$10^{-3}$ & 2 & $2.24 \cdot 10^{-9}$ & 0.970 \\
\hline
$10^{-2}$ & 1000 & $1.00 \cdot 10^{-8}$ & 0.992 \\
\hline
\end{tabular}
\end{center}


\bibliographystyle{unsrt}  
\bibliography{references} 

\end{document}